# Enhanced functions supporting process planning for aircraft structural parts

Ramy HARIK*†, Vincent CAPPONI**, Muriel Lombard*, Gabriel RIS*

*CRAN, Research Center for Automatic Control, Nancy, CNRS UMR 7039
Université Henri Poincaré, Nancy I, Faculté des Sciences
Vandoeuvre-lès-Nancy, BP 239 – 54506, France
Phone: +33-3-83684419, Fax: +33-3-83684437

** Institute of Production and Robotics, Ecole Polytechnique Fédérale de Lausanne
EPFL-STI-IPR-LICP, Station 9, CH-1015 Lausanne, Switzerland
Phone : +41 21 69 35 915, Fax : +41 21 69 33 509

† **Corresponding author:** ramy.harik@cran.uhp-nancy.fr

**Abstract** – Aiming at automating the different trades intervening in mechanical parts' design activities, one finds the weakest link within the interface linking the computer aided-design trade (CAD) and the process planning trade (CAPP) where a huge semantic gap is detected. Works generally tends to optimize the pre-process planning trade (design/CAD) and post-process planning trade (machinning/CAM). So as to bridge in between CAD and CAPP we present at first support functions that helps a process planner setting a process plan for an aircraft structural part. The proposed functions will be forwarded in the same order the planner questions himself during his first analysis. Each function is justified by presenting the need behind. We end up on the benefits of having such an assistance tool as we mention the future perspectives of our works.

## I. INTRODUCTION

Tending towards shortening the production time cycle, manufacturers identified the different intervening trades and studied the areas where the production can be accelerated. We focus our study between the design and manufacturing process. Three different trades intervenes generally, the design where the product specifications are established, the process planning where the process plan is set up and the manufacturing where the milling parameters are selected. Works in between these trades started evolving around each as a stand alone trade.

CAD systems were first revealed in the 60's, however they were drawing oriented without any ability to conduct mechanical calculus in behind. In the late 70's CAD systems performance witnessed a new era and research on features-based systems accelerated whereas in the 90's these systems became an industrial need due to their high performance ability. In parallel, works on CAM systems were being held. But there were no real interfacing in between the two different trades. Works on CAPP systems started around the 80's with the first systems providing the ability to generate process planning being developed in few industrial ventures. While the interface between CAPP and CAM is well integrated (since they are manipulating the same objects), there is a huge semantic loss between the CAD and CAPP trade, where even the support models used by these systems are completely different. Different works were publish around, whether using STEP (a neutral product data exchange system) or machining/design features, still the semantic gap is not fulfilled mainly in areas of interest that have high degree of specificity.

French RNTL project UsiQuick was established to answer this need in the aeronautical field. Analysis conducted showed that two main reasons justify the need of automating the design/process planning interface as well as characterizing CAPP supporting functions. The first being the small production batch when it comes to the aeronautical manufactured parts, the second being the loss of expertise knowledge due to mass retirements of people that first worked in this domain.

Through the different sections of our articles, we will forward a state of the art evolving around the need of automating process planning. Once justified, we shall forward functions proposed to simply the aeronautical process planner task. The functions are advanced in identical order to how a process planner questions the part. Each function presentation is made first by presenting the need and utility behind, followed by the characterization algorithms and application.

## II. STATE OF THE ART

Process planning is a function within manufacturing facilities that selects manufacturing processes and parameters to be used to transform a part from its initial stage to the final stage based on a predefined engineering drawing [Zha, 94]. Many researches were made to automate the process planning trade. [Alt, 89] defined a list of the tasks a complete design to machining software should include. The first task is the transformation of the CAD model into a comprehensible product model for the manufacturing area. Whereas the first studies proposed a complete automation of the process planning trade, research is recently tending towards characterizing support functions to the process planning trade, without eliminating the expert interaction.

[Ans, 94] specified that process planning doesn't have a structured generation method, no proved techniques, no synthesized references, and no norms. It is completely based on the expert's experience and the available methods in the company. [Vil, 99] adds that process planning automation evokes a multitude of opposed criterions and contradictory objects while requiring a huge expertise and knowledge, uneasy to be modeled and coded.

Process planning generation strategies can be classified through two different categories: variant and generative ones.

What we identify as variant process planning is a retrieval of old process plan based on the part's morphology. It usually concerns look-alike parts' domains [Mut, 88 - Cha, 90] such as the hydraulic one. Variant systems identify previously proposed process plans for similar parts and try to adjust them to the new part. This method proved that it is difficult to create and maintain. [Li, 94] presents an industrial venture where after spending 2 years inserting data within the knowledge database, it was still almost impossible to integrate new manufacturing technologies due to time consuming database inquiries. It is to mention that this method is completely useless in the aeronautical field since we rarely find two similar parts.

Generative systems consist in integrating the process planner knowledge within the process planning generation system. They are usually split in between algorithmic methods [Wys, 77 – Che, 92], when the knowledge is hardly coded within the system, and expert systems [Des, 81 – Tsa, 87], when the knowledge is being constructed in a separate database and only called once the specific trigger reason is identified. Recently, Artificial Intelligence systems are being questioned.

One important aspect in generative methods is how to formalize the planner's knowledge from one side, and how to memorize it within the systems, from the other. Usually knowledge in machining is stored by two means [Vil, 03]. The first being production rules: IF {reason} THEN {consequences}. The second by constraints: Geometrical constraints (same setup for two features), topological constraints (feature one processed ahead feature two) and kinematical ones (feature processed using this milling direction). To propose a framework during the reasoning, machining features are widely used.

[Gam, 90] defines machining features as the association of a milling process, a geometrical shape and specifications. [Sha, 95] gave a detailed definition: Machining features are a set of geometrical elements which corresponds to a certain machining process. These elements can be used to generate the appropriate machining means that can help creating the concerned geometry.

Feature modeling is often seen as the answer to support the passage from CAD to CAPP. However these features are not widely used in nowadays systems. This consequences leads to the need of a machining oriented product model able to support the different concerned trades.

A confidential industrial state of the art concludes that through the establishment of certain process planning help functions, programming time would be reduced up to 70%. As the aeronautical parts are produced in very small batches, any reduction of the programming time leads to substantial savings on the total cost of each part.

If we were to conclude the presented state of the art in three points, they would be:

- We renounce the complete automation idea towards the specification of a performing support system,

- The generation of a machining oriented product model is essential to contain the newly generated information,

- Support functions are needed to reduce the enormous time spent on setting up aeronautical parts' process plans.

III. PROCESS PLANNING FUNCTIONS

By the following we shall present some of the functions that answer the first questions a process planner questions when he first sees an aeronautical part (fig 1). Those functions were extracted after extensive studies and interviews with process planners from aircraft companies. The level of details given in some of the presented functions was subject to confidentiality clause, and thus only the main guidelines are presented in this paper. These functions can be generalized to other industries as stand alone functions; however the proposed sequencing is representative of aircraft industries.

The functions are proposed to experimented process planners, presuming that all the identified attributes were calculated and filled in the part's newly generated machining oriented product model.

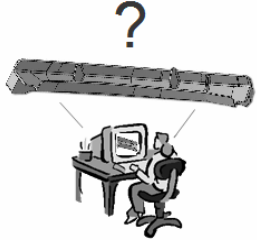

Fig. 1. Scheme representing the aim of our article: To answer the questions a process planner asks when he first sees the part

*A. Dimensions and stock selection*

The first evoked question evolves around the part's dimensions. The planner checks if the main dimensions of the simplest stock around the part (fig.2) remain compatible with the workspace of the available machine-tools in the workshop, else he may have to contact another workshop or outsource the manufacturing of the part. If the part is small considering the average workspace, he may consider machining several parts in the same stock, therefore in the same set-up (fig.3).

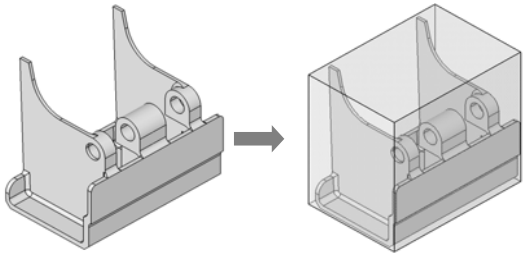

Fig. 2. Minimal stock of the part

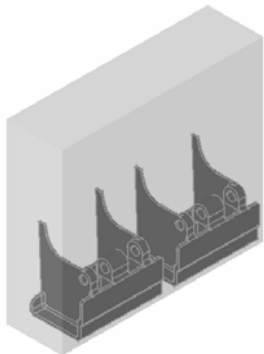

Fig. 3. Multiple parts in the same stock.

The support function for this task consists in providing a list of the compatible machine-tools considering the part dimension. For this rough sorting, the minimal stock (which will not be the real stock) will be computed and compared to the workspace of the available machine-tools. For each machine-tool, the theoretical number of parts machinable on the same set-up can be calculated too. However, multiple parts in the same set-up increases the difficulty of the planning tasks, so next functions proposed in this paper does not tackle the multiple part configuration.

*B. Potential milling tools*

While investigating the part, different generated attributes will be useful for selecting the finishing milling tools. Milling tools are a roller with indented edge or surface, for producing like indentations in metal by rolling pressure, as in turning. Their selection depends heavily on two main criterions. The first being the type of the material: special cutting tools are made depending of the stock material (aluminum, titanium…). The second criterion depends on the part's geometry. Upon the part's analysis we define three main parameters: the cutting length, the tool's diameter and its corner radius. This first selection is made considering standard fillet-end cutter.

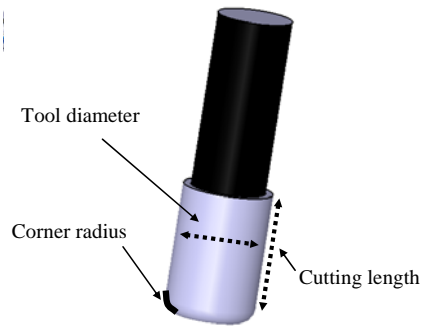

Fig. 4. Geometric characteristics of the Fillet-End Cutter

Different methods are called to determine the three parameters.

The cutting length is selected upon ray launching techniques. Once a milling mode is selected or that the face is characterized as potentially machinable, and after the determination of the accessibility vector (check §3.D and §3.E), we can compute the minimum length of the to-be-removed material.

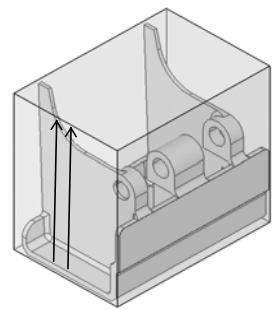

Fig. 5. Ray tracing to determine cutting length

The tool corner radius is computed upon the smallest fillet radius of the part. A fillet is the junction surface often characterized by a cylinder whose radius is less then

4mm or a sweep face whose radius is less then 6mm. usually designers omit placing fillets unless they perform certain functionalities.

The final parameter would be the tool diameter. It is directly related to the selected milling mode. If end milling, then the diameter is selected by offsetting the surface and checking the interactions with it's surrounding. If flank milling, we check up the different adjacent faces and their type (planar, cylindrical, ruled, conical…). And then we select the smallest cylindrical radius to perform the milling.

Having generated a first potential milling tool, we can proceed by iterating and checking the results. If convenient we fix our selection, else we iterate to reach the optimal tool. The cutting length should not be too long else it might induce vibration and deflexion. Once 95% of the face is milled we can consider the tool as satisfactory and proceed what remains with another tool.

*C. Machining difficulties*

To solve machining difficulties, some features have to be finished while proceeding with the roughing. We shall present one of these particular features: "thin features".

"Thin features" are characterized by two planar surfaces separated by a small distance in between them as shown in figure 6.

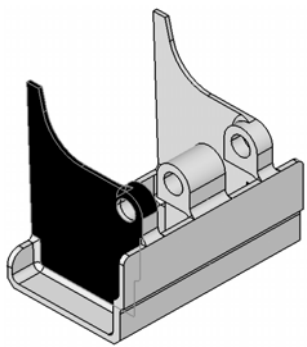

Fig. 6. Thin features

These features can be enclosed in the part (they will be thus characterised as bottom thin features), or half enclosed (and they might be characterized as wing thin features).

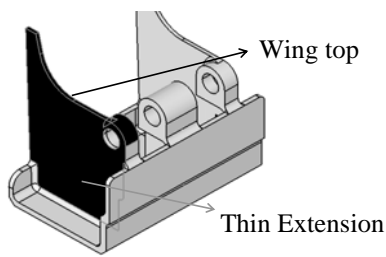

Fig. 7. Thin features nomenclature

The particularity of these features is essentially their milling priority. A wing top should be finished while roughing the part. This is explained by the effort generated while proceeding with the material removal.

If we take the example of the presented mechanical part, the thin extension consists of two faces, one shown on the fig 7, and another internal (from the other side). The external part will be end milled while the inside one will be flank milled due to accessibility reasons. The wing top will also be flank milled. An understanding of the mechanical forces generated while proceeding with the material removal (fig 8) reveals that the forces present on the part might induce a higher part's deformation when not milled in a particular order.

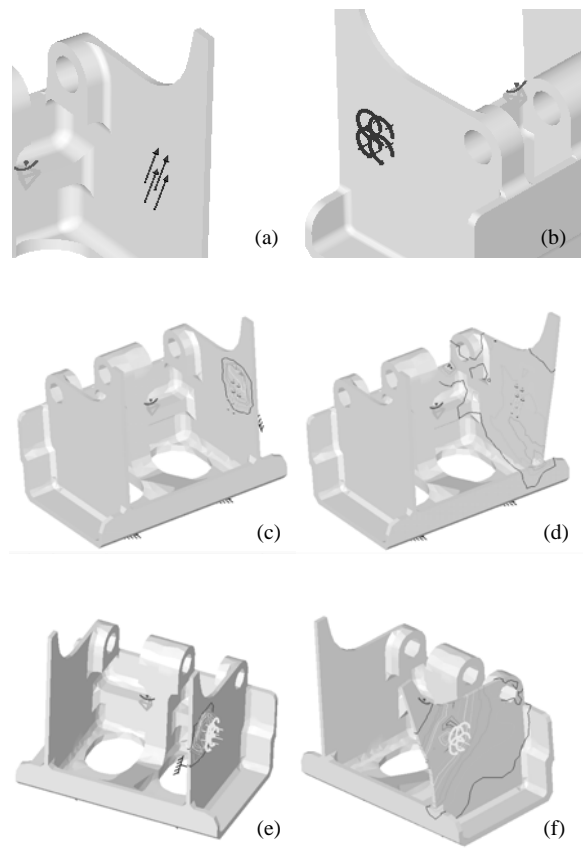

Fig. 8. (a) Flank milling forces simulation, (b) end milling forces simulation, (c) internal flank milling deformations when external end milling is not processed, (d) internal flank milling deformations when external end milling is previously processed, (e) end milling deformation when inside flank milling is not processed, (f) end milling deformations when flank milling is previously processed..

The figure above shows the orientation of the forces, when the milling tools are removing the material. The flank milling of the web top is to be conducted first; hence the force orientation would deform the part. If finished before roughing the thin extensions, the force efforts will be propagated into the to-be-removed material. Then we would flank mill the internal side before end milling the external side. The end milling forces being in the planar section the different resulting stress forces won't lead to the part deformation unlike the internal side flank milling.

We conducted a simple static deformation calculus on the part in order to confirm these assumptions. We consider two sequences of finishing operations: (1) flank milling the inner side before end milling the outer side and (2) flank milling the inner side after end milling the outer side. According to thin walls standard strategies, large finishing allowance (5mm) has been set. Figure 8 illustrates the different mentioned cases and Table 1 presents the maximum deformation values during the machining of the considered wall.

| Sequence | Deformation during flank-milling | Deformation during end-milling |
| --- | --- | --- |
| (1) Flank + End | $5.10^{-5}$ mm (c) | $4.10^{-5}$ mm (f) |
| (2) End + Flank | $4.10^{-2}$ mm (d) | $1.10^{-3}$ mm (e) |

Table 1. Maximum deformation values

By identifying these features we would be generating the first process planning phase within an internal priority ranking to respect.

*In the following two sections (D and E) we will inquire the face's ability to be milled with end and flank milling respectively. The following figure (Fig. 9) represents the origins of this nomenclature. It is due to the used tool section in the part material removal process.*

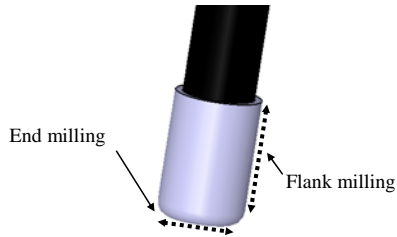

Fig. 9. Different milling modes, flank milling when using the tool's side, end milling when using the tool's end, and simultaneous when using both.

*D. End Milling*

The ability of a part's face to be end milled is studied next. The approach is based on visibility and accessibility issues. [Der, 05] has proposed different solutions based on finite elements and topological operations.

In his topological approach, the following algorithm was proposed, once the face is selected (fig 10.a):
- We construct an infinite extrusion of the studied face (fig 10.b),
- Then we intersect the obtained extrusion with the originating part (fig 10.c),
- The result is then projected on the original face and,
- By eliminating inside parts of the projection (cleaning phase), we obtain the masked zone (fig 10.d).

Through the application of this algorithm on the part's different faces, we will be able to determine which ones can be machined using the end milling mode. And moreover, we are able to determine the exact percentage of the face that will be machined.

Certain process planners will select end milling even if the unmachined area reaches 20% of the part, and then will machine the remaining with either flank or sweep milling.

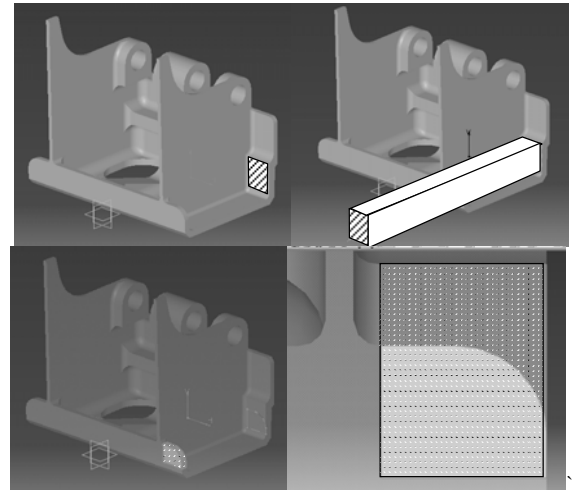

Fig. 10. End milling computation different steps. In the end, we obtain the masked zone (in horizontal lines) and the machinable zone (in vertical lines).

In general, we type the part's face with the end milling mode, when 95% of the face is doable. The remaining 5% can be then hand milled or proceeded with other material removing techniques.

*E. Flank Milling*

In the particular sector of aeronautical structural parts, experts claims that flank milling is the optimal machining mode for the face's quality, cost and milling time. This finding is due to the enclosed geometries that restrict the potential end-milling tools and usually lead to select cutters with small diameter. In this case,, flank milling may be 14 times faster than end milling due to the wider tool-part contact area. At the same time, the shape generated by the flank of a tool is often of finishing quality and having reduced the machining time the cost automatically decreases.

However flank milling's trajectories generation for webs or stiffeners is of a certain complexity, that actually planners prefer end milling on flank milling. Within our PhD studies, we suggested different modules to treat flank milling.

Ruled surfaces can only be milled using the flank milling mode. Ruled surfaces are the result of moving a rule along a certain trajectory. Planar, cylindrical and conical faces are particular ruled ones. The treatment of flank milling is separated between the planar surfaces on one side, and the general ruled surfaces, the cylindrical and the conical on the other. This is mainly due to the infinite number of directions that a planar surface can be machined with while the general ruled surfaces can only be milled following the rule's trajectory.

The study generally consists of identifying the ruled surface, extracting the machining directions [Har (a), 06], and computing the visible zone [Har (b), 06] to validate the selected machining directions.

The machining directions for planar surfaces are based on industrial knowledge extraction computations. The method relies on three aspects: the face' edges, the face' vertex transitions and the backward edges. The first being a local treatment depending on the edge' different parameters (sharpness, angles, adjacency…). For a certain value of the different parameters a certain machining direction will be proposed. The second relying on the point transitions (sharpness of the bounding edges), in fact we try to relay the most constraining vertexes with the least constraining ones (which might eventually constitutes accessible areas for the face). The last would be a special treatment for the different edges that are not accessible through their knowledge proposed machining direction.

For other ruled surfaces the machining directions are constituted of the set of rules generating the surface.

Once the machining directions are proposed we call on different modules to validate them. These methods rely heavily on topological computations to extract the visible and masked zones. Having multiple machining directions attached to each face we compute each direction's accessibility on the part. Then, we intersect all the results obtained on a single face to generate the final result: the global visible area and the G-Zones [Har (b), 06].

*F. Toward a support for positioning the part on the machine-tool.*

The function we present below is very dedicated to the machining of aircraft structural parts, meaning that it is the first step to support a specific planning activity, non-usual in programming for other components such as automotive or mould parts [Cap, 06]. Indeed, structural aircraft parts are made from slab stock in two setups (front-side/back-side) on a five-axis machine tool with no intervening operations. The fixture elements are included in the stock and the workpiece remains connected to the stock by tabs at the end of the machining. To post-finish the part, the tabs are manually cut by an operator. As the initial orientation of the part in the stock is not specified by the designer, the planner could set it in order to ease the machining. For the two set-ups, the slab stock will be stick on the table of the machine-tool; therefore the stock orientation question is totally related to the positioning of the part on the machine-tool.

At the very beginning of the set-up planning, the expert questions itself how to set the stock ensuring that all machining directions of the parts are reachable within the two opposite set-ups. By his experience, the planner usually identifies the "bottom of the part" and set the great plane of the stock parallel to this "bottom". Then, he checks that all the machining directions are reachable respecting the rotational degree of freedom of the selected machine-tool. If that is not the case, he has to consider another stock orientation or additional set-up.
As the identification of the "bottom of the part" relies on cognitive interpretation and implicit information that cannot be included in the manufacturing product model, we propose a heuristic method based on the previously calculated criterions:

- We select the plane features that were identified as "can be machined with an end milling mode",

- Within this selection, we cluster in the same group the planes that are parallels , and calculate the total area of each group (sum of the area of all the clustered planes),

- We select the group with the biggest area and set the main plane of the stock parallel to a plane of the group.

To summarize, we propose a plane from the Manufacturing Product Model that implicitly define the orientation of the part on the machine-tool. This function relies on the previously presented machinability functions (end-milling or flank-milling ability) to match with the planner way of thinking.

Finding this plane automatically will allow displaying directly the part on the machine-tool model (two degrees of freedom are now fixed), helping the planner to consider accessibility constraints. Indeed, the next step for the planner is to set the position of the part ensuring that the limited motions of rotational axes do not forbid the access to any of the selected machining directions.

[Cap, 05] proposed several other methods to set the stock orientation of for aircraft structural parts, one of which was to compute the stock of minimal volume. They also proposed a visibility-based method and an algorithm for

positioning the part on the machine-tool and accessibility checking [Cap, 04]. Their usage is currently questioned in industry and will serve as a future function proposition for our process planning generation help system.

*G. Sharpness reduction*

The final function we shall present is relevant to assembly workers safety. Aeronautic parts have often sharp boundaries which might lead to hand cutting. Already identified in the part characterization, the algorithm combines the edge's length and sharpness. If the safety condition is unavailable (i.e. over 140 mm of length, with an open sharpness) we process to sharpness reduction on the selected edge. The reduction might be through a particular industrial method, or through local milling treatments on the edge.

## IV. APPLICATION

The presented approach was developed thanks to the open architecture of the CATIA® V5 PLM software with CAA (Component Application Architecture). The development consisted of different phases to reach the actual status.

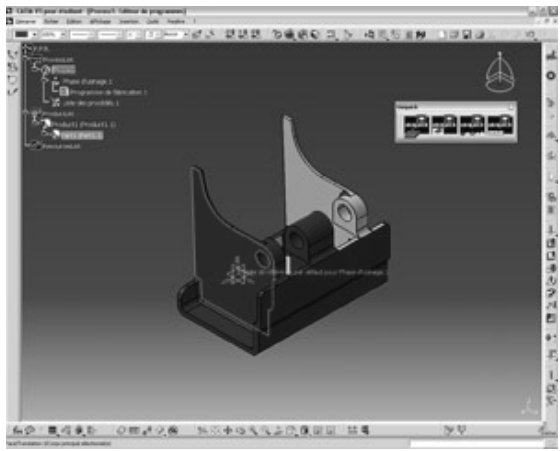

Fig. 10. The functions development completely integrated within the CAD/CAM to PLM software.

The presented result is the effort (fig 10) conducted within the UsiQuick (French RNTL) project. Four laboratories (besides ours – the CRAN) participated in the effort (IRCCyN – Nantes, LGIPM – Metz, L3S – Grenoble and LURPA – Cachan). Two Industrials took part of the venture too (DASSAULT AVIATION and DASSAULT SYSTÈMES) and an expertise center (CETIM).

It is currently applied within a specific IT Tool with perspectives of accepting neutral CAD formats. The actual status of the software nowadays proposes some of the solutions to the prescribed functions, while the remaining is in current application.

Figure 11, represents of the results actually obtained by applying the flank milling detection (function E) on a certain mechanical part. The presented arrows are directed in opposite terms to how the milling tool will approach the part.

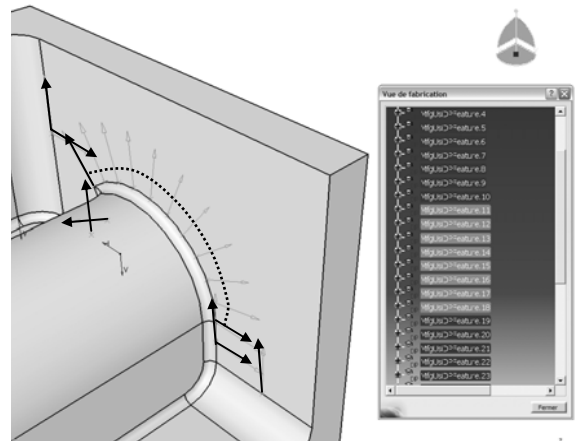

Fig. 11. Planar machinning directions resulting from the application of the E (Flank Milling) function. [Har (b), 06].

## V. CONCLUSION

The aim of our article was to present functions a process planner requires to facilitate his time-consuming effort. At first, we proved through our state of the art, that providing performing help functions is healthier then the approach tending to automate the process planning activity. Then, we presented the functions: Dimensions and stock selection, Potential milling tools, Machining Difficulties, Particular features, End milling, Flank milling, Positioning and Sharpness reduction. Each of the functions was at first justified then parts of the algorithm were presented. We ended the article with the presentation of the application in a CAD/CAM to PLM software.

The presented functions can reduce to over 50% the actual time a process planner requires to study a mechanical part. This percentage was established due to interactions with process planners in the aircraft industries. Future perspectives comprehend the characterization of other software functions. Our main effort is actually concentrated on linking the different results of these functions to assure coherence with the different outputs the actual assistance module generates.